# Stability of planetary systems within the S-star cluster: the Solar system analogues

N. Davari[1]⋆, R. Capuzzo–Dolcetta[1], and R. Spurzem[2,3,4]

[1]*Dep. of Physics, Sapienza, Univ. of Rome, P.le A. Moro 5, Rome, Italy*
[2]*Zentrum für Astronomie der Universität Heidelberg, Astronomisches Rechen–Institut, Mönchhofstr. 12-14, 69120 Heidelberg, Germany*
[3]*Kavli Institute for Astronomy and Astrophysics at Peking University, 5 Yiheyuan Rd., Haidian District, 100871, Beijing, China*
[4]*National Astronomical Observatories and Key Laboratory of Computational Astrophysics, Chinese Academy of Sciences, 20A Datun Rd., Chaoyang District, 100101, Beijing, China*

**ABSTRACT**
A dynamically relaxed dense cluster comprised of about 40 stars (the so-called S-stars) inhabits the central region of the Galaxy. Their stars revolve around the Sgr A$^*$ massive object. To understand the dynamical evolution of planetary systems in a particular environment like that around Sgr A$^*$, we carry out direct N–body simulations of planetary systems embedded in the S–star cluster. In this work, we investigate the short-term stability of the planets orbiting around S-stars after their close interactions with the central massive black hole of our galaxy. We find that planetary systems go through encounters with the SMBH and the nearby stars. We determine the frequency and the strength of planetary systems' encounters with the nearby stars as these encounters remarkably increase for systems assigned to S-stars closer to the SMBH. The SMBH severely destabilizes the planetary systems, though we noted that the small oscillations in the mutual eccentricity and inclination of the planetary system could be caused by the planet–planet coupling and the near-resonance effect between the two planets. We obtain estimates of the fraction of survivor planets (∼ 51%), and find that planets stripped from their hosting star are generally captured on close orbits around Sgr A$^*$. We notify while gas giants are tidally disrupted, terrestrial planets do not. We estimate that Sgr A$^*$ flares can be due to the tidal disruption events of *starless* giant planets.

## 1 INTRODUCTION

Most stars form as part of a stellar group and are mostly surrounded by a disc of dust and gas from which potentially a planetary system might form (Bhandare & Pfalzner 2019). Thereby star formation is inextricably linked to planet formation (e.g. Parker 2020). Circumstellar discs are a natural by-product of the star formation process and most stars, independent of mass, are born with the ability to form planetary systems (Lada & Lada 2003; Fedele et al. 2010; Vincke & Pfalzner 2018; Cai et al. 2019). The prevalence of planets in the universe is now broadly acknowledged. To date, more than 4000 extra-solar planets have been detected[1]. Surveys for planets orbiting solar–like stars have demonstrated that planets are common and may even outnumber stars in the Milky Way galaxy (e.g. Cassan et al. 2012; Mulders et al. 2018; Hoppe et al. 2020). Cassan et al. (2012) found that on average every star hosts one planet thus planets around stars in our Galaxy seems to be the rule rather than the exception. Recently, there have been speculations on whether stars at the Galactic Centre (GC) could also possess planetary companions. The GC is the Milky Way's most important star forming region, a uniquely accessible laboratory for exploring the interactions between a supermassive black hole (SMBH) and its stellar environment.

The young stellar clusters in the GC, the Arches, Quintuplet and the Central Parsec clusters, are located within ∼ 30 pc in the projection of the radio source Sgr A$^*$ possible sites of the remarkable burst of star formation in the GC (Figer et al. 2004, 1999; Gallego-Calvente et al. 2021). The Arches cluster is a massive, dense young cluster of stars (Figer et al. 2002; Krivonos et al. 2014) that extraordinarily could survive despite the closeness to the GC manifesting a significant population of circumstellar discs. The detection of strong disc emission from the Arches stars is surprising in view of the high mass of the B–type main sequence host stars of the discs and the intense starburst environment (Stolte et al. 2010).

Another densely packed group of somewhat fainter ($m_K$ = 14 − 17) and probably lower mass stars, the so-called "S-star" cluster, lies within a distance of ∼ 0.04 pc from the radio source Sgr A$^*$

⋆ E-mail: nazanin.davari@uniroma1.it
[1] According to the NASA Exoplanet Archive

(e.g. Ghez et al. 2003; Genzel et al. 2003). Like any other star cluster in the Milky Way the S-star cluster could be a planetbirth region. Over the last 2.7 decades radio and infrared observations of the central arcsecond of the Galaxy have provided detailed information of a population of both young and old stars in the S-star cluster, in the magnitude range $m_K = 14 - 17$ (e.g. Schödel et al. 2002; Ghez et al. 2003; Gillessen et al. 2009, 2017; Habibi et al. 2017, 2019). The mass of compact radio source Sgr A*, associated with the SMBH of the Milky Way, has been measured with high accuracy through the observation of the orbits of individual S–stars, in particular the highly eccentric orbit of S2. The highly-concentrated mass responsible for the orbital motion of the S-stars is $\sim 4.3 \times 10^6$ $M_\odot$ at the distance $\sim 8.33$ kpc from us (e.g. Schödel et al. 2002; Ghez et al. 2003; Genzel et al. 2003). S2 orbiting the compact radio source Sgr A* is a precision examine of the gravitational field around the closest SMBH candidate (Gravity Collaboration et al. 2020).

Recent spectroscopic analysis within the $0.04 \times 0.04$ $pc^2$ of the GC by Habibi et al. (2017) reveals the estimated age of < 15 Myr and a spectral type of B for the early–type stars. An age of $\sim$ 6.6 Myr is derived for the brightest star of this group, S2, which is compatible with the age of the clockwise rotating young stellar disc in the GC favor a scenario in which these early–type stars formed in a local disc rather than the binary-disruption scenario throughout a longer period of time (Habibi et al. 2017). Whereas in a dynamically relaxed cluster around an SMBH, such as the S–star cluster, a dense stellar cusp of late–type stars is expected to form. Ghez et al. (2003); Habibi et al. (2019) found a cusp in the surface number density of the spectroscopically identified old (>3 Gyr) giants population within the central arcsecond of the Galaxy.

In total, Gillessen et al. (2017, 2009) have determined orbits for 40 stars of the S–star cluster so far, a sample which consists of 32 young stars of spectral type B0-B3V with inferred masses lie within $8-14M_\odot$ (Habibi et al. 2017) plus 8 late–type stars (G, K, and M type) in the mass range $0.5-2$ $M_\odot$ (Habibi et al. 2019). Among those 32 early–type stars, 24 stars are with randomly oriented orbits and a thermal eccentricity distribution while the other 8 are members of the clockwise (CW) disc of young stars with lower eccentricity orbits (Gillessen et al. 2017).

Very young ages are somewhat mysterious because strong tidal forces of the SMBH should inhibit local star formation (Morris 1993). For this reason, the hypothesis of a migration, due to binary disruption, is more likely to explain their presence in the vicinity of Sgr A* (Hills 1988). Therefore, it sounds reasonable that the S-stars formed elsewhere and migrated to their current locations (Antonini 2013).

In addition, the central 0.1 parsecs of the Galaxy host a population of dust-enshrouded objects orbiting the Galactic black hole. Gillessen et al. (2012) reported the presence of a dense gas cloud approximately three times the mass of Earth that is falling into the accretion zone of Sgr A*. A faint dusty object, nicknamed G2 by Burkert et al. (2012), orbiting the SMBH on a highly eccentric orbit with a pericenter radius of roughly $\sim$ 2000 Schwarzschild radii from the central massive black hole Sgr A* (Pfuhl et al. 2015). Pfuhl et al. (2015) assessed the semi-major axis and eccentricity of $a = 0.042 \pm 0.01$ pc and $e = 0.98 \pm 0.007$ for the G2 cloud. Another dusty, ionized gas cloud of moderate mass, analogous to G2, is reported in the vicinity of Sgr A* (Ghez et al. 2005; Clénet et al. 2005; Pfuhl et al. 2015). This object, the G1 cloud, is on a smaller orbit ($\sim 0.01$ pc) with lower eccentricity ($\sim 0.86$) compared to G2. However, both G1 and G2 have remained intact with no apparent sign of disruption after passing through pericentre (Haggard 2014; Pfuhl et al. 2015; Mapelli & Ripamonti 2015). Ciurlo et al. (2020) report observations of four additional G-like objects, all lying within 0.04 parsecs of the SMBH Sgr A* and forming a class that is probably unique to this environment. The four objects show many properties in common with G1 and G2 (compact Br$\gamma$ emission lines and coherent orbital motion) and therefore named G3[2], G4, G5 and G6 [3]. Hence, the G–objects could have a joint origin. While G2 is called "gas cloud" in the first paper by Gillessen et al. (2012) other models proposed that a faint and undetected star is at the origin of G2 (Gillessen et al. 2013). Murray-Clay & Loeb (2012) proposed that G2 might be the evaporating protoplanetary disc around a faint (young) star. Mapelli & Ripamonti (2015) discuss the possible capture by Sgr A* of planets or planetary embryos orbiting the young stars in the innermost parsec of our galaxy and bring them onto nearly radial orbits since several lines of evidence signify on-going star formation within two parsecs of Sgr A* suggesting a population of photoevaporative protoplanetary discs associated with newly formed low mass stars (Yusef-Zadeh et al. 2015a,b, 2017).To investigate whether or not planets could exist in the vicinity of Sgr A*, we can suppose that stars in the Galactic Centre (GC) already harbour planetary systems of different masses and semi-major axes and explore if there is any peculiarity from these planets that is detectable. In Capuzzo-Dolcetta & Davari (2020), we modeled the dynamics of binary stars where each star has one planet orbiting around it as they move toward the Sgr A* massive object. This study can be considered also as a profound quantitative generalization of the Hills (1988) scenario to the presence of planets around a binary star. In Capuzzo-Dolcetta & Davari (2020) we showed the intriguing possibility of the existence of hyper-velocity stars keeping planets around, as well as the production of solitary hyper-velocity planets and, also, the existence of planets bound to stars which have turned to S-stars revolving around the central massive object. Trani et al. (2016) investigate the dynamics of planets and protoplanets near the SMBH in the GC. They found that tidally captured planets which were initially bound to CW disc stars remain in the CW disc. On the contrary, their simulations for the S-star cluster illustrate that planets initially bound to S-stars are captured by the SMBH on highly eccentric orbits, matching the semi-major axis and eccentricity of the G1 and G2 clouds.

Many studies have examined the survivability and dynamical evolution of multi-planet systems in star clusters (e.g., Spurzem et al. 2009; Cai et al. 2017; Flammini Dotti et al. 2018, 2019; Li et al. 2019; Stock et al. 2020). Using both direct N-body and hybrid Monte Carlo simulations, Spurzem et al. (2009) investigate the dynamics of single-planetary systems in star clusters and provide a self-consistent model of the frequency of planetary systems in the long-lived open and globular clusters due to encounters of planetary systems with passing stars. Sufficiently close stellar encounters, which are likely to happen in dense clusters, can induce high planets' orbital eccentricity causing dynamical instability, especially in strongly packed systems. The detached planets generally remain in their host clusters as free-floaters. The discovery of freely floating low-mass objects in young stellar clusters such as $\sigma$ Orionis can be considered a supporting piece of evidence for this scenario. However, Spurzem et al. (2009) noted that short-period planet orbits are more difficult to be destroyed by stellar encounters which can excite only modest eccentricities, such that orbital decay, tidal inflation, and even disruption of the close-in planets are the subsequent effective tidal dissipation phenomena. To understand better the role of stellar encounters in the destabilization of planetary systems, Cai et al. (2017) performed *N*-body integrations for four different planetary system models in three different star cluster environments (*N* = 2k, 8k and 32k, respectively). Cai et al. (2017) considered an ensemble of initially identical planetary systems (consist of either five $M_J$ or five $M_⊕$ in initially circular and coplanar orbits) around solar-type stars and followed their evolution for 50 Myr. Cai et al. (2017) showed that planet-planet interaction induced by stellar fly-by perturbations can cause the fragility of planetary systems inducing a significant low-mass planet ejection. Albeit, from an observational point of view, Cai et al. (2017) results anticipate higher planet detection rates in the young low-mass star clusters. In order to understand the implications of a massive object contribution for the evolution and disruption rates of planetary systems, Flammini Dotti et al. (2020), investigated the influence of a central intermediate-mass black hole (IMBH) in star clusters. For a star cluster with a 100 or 200 $M_⊙$ IMBH, their findings show that the IMBH has the most prominent effect in the rate at which planets are expelled from their host star. Although the presence of IMBHs in the centres of star clusters has not yet been determined, their study highlights the importance of a central massive object when considering planetary systems in star clusters.

---

[2] G3 was previously identified as D2 (Eckart et al. 2013; Sitarski 2016)
[3] Although, G6 has been independently examined, and interpreted as a bow shock source instead of a G object (Peißker et al. 2019)

In the above sketched framework, we intend to numerically explore the short-term stability of putative planetary systems around stars in the S-star cluster after close interaction with the central black hole of our Galaxy, and inspect their dynamics and stability in the hostile environment of the Galactic Centre (GC).

In section 2, we describe the setup of our numerical simulations. In section 3 we provide a description and analysis of the results; first, in 3.1, we focus on the statistics of encounters between a planetary system and the neighbouring stars to determine the frequency and the strength of these encounters. In 3.2, we pay special attention to the role of the Kozai–Lidov oscillations in the destabilisation of the planetary systems and, differently from other studies, explore small oscillations in the mutual orbital parameters of the planetary systems could be caused by the planet–planet coupling and the near-resonance effect between the two planets rather than the Kozai–Lidov effect. In 3.3 the tidal disruption events of *starless* giant planets are studied. Finally, section 4 contains a summary & discussion of the findings of the work.

## 2 NUMERICAL SIMULATIONS

The advancement of both software and hardware has enabled to employ direct *N*-body simulations to follow the evolution of star clusters. Direct simulations are the most reliable because of their accuracy and few assumptions adopted. Actually, the exploration of the environment of galactic nuclei has fruited significantly of the use of *N*-body simulations. However, it is worth noting that the presence of an SMBH in a galactic nucleus can cause particular difficulties to straight *N*-body integration. Close approach of 'stars to the SMBH leads to an enormous acceleration making a high–accuracy regularization treatment essential to give valid results.

### 2.1 Initial Conditions

As of today, the number of stellar orbits around Sgr A* which have been determined amounts to about 40; 8 stars of this group belong to the CW disc (Gillessen et al. 2017). They constitute the so called S-star cluster.

In this work, we employ direct N–body simulations to investigate the dynamical stability of planets attached to the 40 innermost stars of the S–star cluster orbiting around the (supposedly) non-spinning Sgr A* SMBH. The orbital parameters of S-stars are taken from data of Gillessen et al. (2009, 2017) as initial conditions. As better explained in subsect. 2.3, the simulations are carried out using a modified version of AR-CHAIN (Mikkola & Tanikawa 1999a,c; Mikkola & Merritt 2008), the ARGDF code (Arca-Sedda & Capuzzo-Dolcetta 2019), an *N*–body code with post-Newtonian corrections up to order 3.5 (3.5PN) that treats the equations of motion by regularization. Furthermore, the ARGDF code authorises to take into account the analytical external matter distributions and treat their dynamical friction.

The mass of Sgr A*, sited in the origin of the coordinate frame, is set to 4.3 ×$10^6$ $M_⊙$. Assuming for the Sun a distance of 8.33 kpc to Sgr A* (Gillessen et al. 2009, 2017), 1 arcsec corresponds to 40.4 mpc (mpc=0.001 pc). The S–star cluster is composed by a population of both early–type (main-sequence) stars and late–type (post main-sequence) giants (Ghez et al. 2003). Habibi et al. (2017) derived the mass range, age, effective temperature, rotational velocity and surface gravity of the main sequence early–B stars of the S cluster and Habibi et al. (2019) identified a cusp of late–type stars around Sgr A* SMBH. In Habibi et al. (2017) the masses of 8 early–type stars in the S–star group have been derived precisely. We apply their estimation in our simulations for the mass of these stars and the other early–type stars with identical magnitude ($m_K$). To assess the mass of the remaining early–type stars (the fainter ones), we suppose a randomly selected mass within 8 − 14 $M_⊙$ as derived in Habibi et al. (2017). In addition, for the old giant population (∼ few Gyr) of this ensemble, the inferred mass lies within 0.5 − 2 $M_⊙$ as in Habibi et al. (2019).

S–stars are cramped around the radio source Sgr A* with the semi-major axes, *a*, in the range ∼ 0.005–0.05 pc. Some of the S-stars approach very closely to Sgr A*, where the relativistic effects could be important to define their orbits. To check the role of the PN terms for the planetary systems orbiting the S-stars, we compared simulations taking into account the post-Newtonian corrections up to order 2.5 for the equations of motion to the purely Newtonian (0PN) integrations. Noticeably, the lowest-order corrections to the Newtonian equations of motion have amplitudes of order $\mathcal{P}^{-1}$, where $\mathcal{P}$ is the "penetration parameter" (Merritt 2013) defined as

$$\mathcal{P} \equiv 2\frac{(1-e)(1+e)a}{r_S} = (1+e)\frac{r_p}{r_S}, \quad (1)$$

where *a* and *e* are the semimajor axis and eccentricity, $r_p = (1 − e)a$ is the distance of closest approach of the S-star to the SMBH and $r_S \equiv 2\frac{GM_\bullet}{c^2}$ is its Schwarzschild's radius. The angular precession per orbital period due to the Schwarzschild contribution to the metric, $A_s$, is (Merritt 2013) $A_s = 6\pi\mathcal{P}^{-1}$. The S-stars with smaller pericenter distances to Sgr A* (for example S2, S14, S31, S38 and S55) have values of $\mathcal{P}$ and $A_s$ which are significant. We noticed that the planetary systems of these stars are more vulnerable. Actually, none of the mentioned stars could preserve intact their planetary systems.

In order to examine the survival chances of unequal–mass planetary systems, we attach planetary systems similar to our Solar planetary system to each of the stars in the S-star cluster. The initial values of the orbital parameters (eccentricities and semi-major axes) for the planets are chosen as those of seven (but Neptune) planets of our Solar system. For the late-type S-stars, the planetary system is composed of the set of 7 planets; Mercury, Venus, Earth, Mars, Jupiter, Saturn and Uranus. The exclusion of Neptune–like planets is because the planet is with the widest orbit ($\sim 30$ AU) in the Solar System. We intend to distribute planets from 0.1–20 semi-major axes based on the Hill radius:

$$r_H \approx a_* \left(\frac{m_*}{3M_{SMBH}}\right)^{1/3} \quad (2)$$

where $m_*$ is the total mass of the star-planet system, $M_{SMBH}$ is the SMBH mass and $a_*$ is the semi-major axis of the star around the SMBH. The above equation gives the $r_H \lessapprox 20$ AU for S-stars with $0.004 \leqslant a_* \leqslant 0.014$ pc while gives $r_H \leqslant 300$ AU for the S-stars with $a_* \geqslant 0.02$. Therefore, the planetary distributions would be very different based on the $a_*$ of S-stars. To have a comparable distribution of planets around S-stars (resembling our Solar system), we generalized the planet's semi-major axis range within 0.1–20 AU on all the S-stars, thus Neptune was excluded.

For the early-type S-stars in our simulations, the considered planetary system is composed of the 3 giants; Jupiter, Saturn and Uranus. Assignment of different planet configurations to early- and late-type stars in the simulations is over two reasons;

(i) Essentially, we have two types of stars with two different mass ranges and it would not be unusual to treat them differently. Nevertheless, Kepler observations have established that planets in short period orbits have a higher occurrence rate around low-mass stars than around solar type stars (e.g., Mulders et al. 2015), so it would not be surprising to find a host mass dependence on the planet orbital width occurrence rate.

(ii) Our initial conditions are limited by a reasonable amount of time with acceptable accuracy. We are therefore restricted in the number of bodies and integration time and we have to rule out terrestrial planets associated with the early–type stars which we found out their presence wouldn't have much influence in stabilization/destabilization of the planetary system.

Additionally, the initial inclination of the planetary systems orbits relative to the host star orbital plane ($i_{in}$) is varied in discrete steps 10° from 0° to 180°. Note that $i_{in} = 0°$ means planets moving on the same plane of the star orbit around the SMBH with same orientation of angular momentum, while $i_{in} = 180°$ corresponds to opposite orientation of the angular momentum.

The S-star with the longest revolution period ($\sim 3580$ yr) is S85, so that we decide to extend the integration time up to 4000 yr to chase at least one of its full revolution around Sgr A*. Each of our runs consists of a total of 193 bodies (stars+planets+SMBH) with a mass ratio up to $\sim 10^{13}$. The average computational time for each run is $\sim$ 48-72 clock-time hours; the total number of simulations is 76. Table 1 shows the assumed parameters of our runs.

## 2.2 Methodology

Our goal is following the orbital evolution of planetary systems around each S–star subjected also to the influence of the perturbation from the Sgr A* black hole, aiming, first of all, at the determination of the survival rate of these planetary systems. We implement a Python driver to generate initial conditions for the whole set of

**Table 1.** From left: the star denomination, the assumed mass or mass range (values for S1,S2, S4, S6, S8, S9 and S12 taken from Table 2 in Habibi et al. (2019)), the spectral type (SP, defined as early (=e) or late (=l)) and the K-band magnitude ($m_K$) both from (Gillessen et al. 2009, 2017; Habibi et al. 2017, 2019), the number of planets ($N_P$) initially around the star. The symbol '*' indicates the stars which belong to the "clockwise disc" and the symbol '+' denotes stars are selected from Gillessen et al. (2009).

| Star | Mass/M$_\odot$ | SP | $m_K$ | $N_P$ |
|---|---|---|---|---|
| S1 | 12.40 | e | 14.8 | 3 |
| S2 | 13.60 | e | 14.1 | 3 |
| S4 | 12.20 | e | 14.6 | 3 |
| S6 | 9.20 | e | 15.4 | 3 |
| S8 | 13.20 | e | 14.5 | 3 |
| S9 | 8.20 | e | 15.1 | 3 |
| S12 | 7.60 | e | 15.5 | 3 |
| S13 | as S12 | e | 15.8 | 3 |
| S14 | as S12 | e | 15.7 | 3 |
| S17 | 0.5 – 2 | l | 15.3 | 7 |
| S18 | 8 – 14 | e | 16.7 | 3 |
| S19 | 8 – 14 | e | 16.0 | 3 |
| S21 | 0.5 – 2 | l | 16.9 | 7 |
| S22 | 8 – 14 | e | 16.6 | 3 |
| S23 | 8 – 14 | e | 17.8 | 3 |
| S24 | 0.5 – 2 | l | 15.6 | 7 |
| S29 | 8 – 14 | e | 16.7 | 3 |
| S31 | as S12 | e | 15.7 | 3 |
| S33 | 8 – 14 | e | 16.0 | 3 |
| S38 | 0.5 – 2 | l | 17.0 | 7 |
| S42 | 8 – 14 | e | 17.5 | 3 |
| S54 | 8 – 14 | e | 17.5 | 3 |
| S55 | 8 – 14 | e | 17.5 | 3 |
| S60 | 8 – 14 | e | 16.3 | 3 |
| S66* | as S1 | e | 14.8 | 3 |
| S67* | 8 – 14 | e | 12.1 | 3 |
| S71 | 8 – 14 | e | 16.1 | 3 |
| S83* | as S2 | e | 13.6 | 3 |
| S85 | 0.5 – 2 | l | 15.6 | 7 |
| S87* | as S2 | e | 13.6 | 3 |
| S89 | 0.5 – 2 | l | 15.3 | 7 |
| S91* | 8 – 14 | e | 12.2 | 3 |
| S96* | 8 – 14 | e | 10.0 | 3 |
| S97* | 8 – 14 | e | 10.3 | 3 |
| S145 | 0.5 – 2 | l | 17.5 | 7 |
| S175 | 8 – 14 | e | 17.5 | 3 |
| R34 | as S2 | e | 14.0 | 3 |
| R44* | as S2 | e | 14.0 | 3 |
| S5+ | as S9 | e | 15.2 | 3 |
| S27+ | 0.5 – 2 | l | 15.6 | 7 |

moving objects and to launch the ARGDFcode to integrate the system over time. The procedure is as follows:

(i) compute positions and velocities of the S–stars from the given set of their orbital elements (a, e, i, Ω, ω, P) as from Gillessen et al. (2009, 2017);
(ii) compute positions and velocities of the planets utilising orbital parameters of their host star (Ω, ω, P), except that the semi-major axes and eccentricities of planets are appended to the ones of each host star and the inclination (i) of planets' orbits with respect to the parent star's orbit rises at 0°–180° with increments 10°.
(iii) as illustrated in Table 1, we need to generate different sets of initial condition for various mass ranges of stars.
(iv) integrate the motion of the stars and planets forward up to 4000 yr using ARGDFfor each set of initial conditions.

### 2.3 The Numerical Code

The `AR-CHAIN` code is based on an accurate direct N–body algorithm suitable to explore the non–relativistic or relativistic (post-Newtonian) dynamics of a few–body system also in presence of a compact massive object such as an SMBH. A relevant characteristic of the `AR-CHAIN` code is that it works at high accuracy at arbitrary mass ratios. The code integrates the equations of motion of the $N$–body systems and uses the so called algorithmic regularization (Mikkola & Tanikawa 1999a,b,c), based on a chain configuration (Mikkola & Aarseth 1993) and on the time transformation leapfrog (Mikkola & Aarseth 2002) to handle with the problem of very large mass ratios. Our simulations have been performed by mean of a modified version of `AR-CHAIN` called `ARGDF` (Arca-Sedda & Capuzzo-Dolcetta 2019), which takes into account both the gravitational field of the galaxy, treated either as a Dehnen (1993) or a Plummer (1911) analytic model, and the stellar dynamical friction effect by mean of an extension of the classical Chandrasekhar (1943) theory. Velocity-dependent forces (like dynamical friction braking) are included via a generalized midpoint method based on the leapfrog algorithm, perfectly suited as in the case of the GC stars around the SMBH (Mikkola & Merritt 2006, 2008). As a matter of fact, we checked that the inclusion of an external potential given by a regular distribution of matter and its dynamical friction effect has a negligible relevance on the dynamics of the set of objects under study here. To date, results for the accurate orbital motion through 3.5 post–Newtonian (3.5PN) order are available (Mora & Will 2004; Jaranowski & Schäfer 1998; Pati & Will 2002). The version of the `AR-CHAIN` code that we employed for our integration incorporates post–Newtonian force perturbations up to 3.5PN (Mora & Will 2004). The 1PN and 2PN corrections are non dissipative and responsible for the pericentre shift (Kupi et al. 2006; Amaro-Seoane 2018) while 2.5PN terms (radiation reaction terms, $O(c^{-5})$ are dissipative and accounts for energy loss via gravitation waves (Blanchet 2014). For our simulations we considered sufficient the inclusion of PN treatment up to order 2.5 included. Relativistic flattening plays no role through 3PN order (Pati & Will 2002) and the first post–Newtonian correction to radiation reaction appears at 3.5 order (Galley et al. 2012) which are omitted in our computations.

## 3 RESULTS

### 3.1 Statistics of Planetary Systems Encounters

Over the 4000 yr of our simulations, the planetary systems are disturbed because of several encounters with the SMBH (with a frequency dictated by the orbital period of the host star) and a significant number of close approaches with the neighbouring stars.

To study the statistics of encounters between a neighbouring star and a planetary system, we quantify the strength of each encounter with dimensionless parameters $K$ and $V_{inf}$, modelling the encounter between the planet and its host star and a third single star as an external perturber with a three–body scattering event inspired by Heggie (2006); Spurzem et al. (2009); Cai et al. (2017). The essential parameters are the relative speed of the third body (passing single star with mass $m_3$) and the star–planet system (of masses $m_1$ and $m_2$), when far apart, $V_{inf}$, and the distance of closest approach between them, $r_p$. The parameter $K$ measures ratio of the timescales involved, defined as

$$K = \sqrt{\frac{2M_{12}}{M_{123}}\left(\frac{r_p}{a}\right)^3} \qquad (3)$$

where a is the semimajor axis of the perturbed planetary orbit (the outermost planets), the mass of the star–planet is $M_{12} = m_1 + m_2$, and $M_{123} = M_{12} + m_3$. In our simulations, if $m_1 = m_3$, since $m_2 \ll m_1$, we take $K \propto (r_p/a)^{3/2}$. Their relative speed at close approach, by a Keplerian approximation, is (Heggie 2006)

$$V^2 = V_{inf}^2 + \frac{2GM_{123}}{r_p}, \qquad (4)$$

while the eccentricity of their relative motion is

$$e' = 1 + \frac{r_p V_{inf}^2}{GM_{123}}. \qquad (5)$$

The frequencies of encounters for planetary systems of the stars S1, S9 and S54 (from top to bottom) are shown in Figure 1. Analogous to figure 1 of Heggie (2006); Spurzem et al. (2009) and figure 2 of Cai et al. (2017), Figure 1 illuminates the distance of pericenter for the passing star (scaled to the semimajor axes of the outermost planet) and the quantity $V_{inf}$ (the relative speed of the perturber with respect to the host star, scaled to the average orbital speed of the outermost planet). The distinctions which are essential in the analytical interpretation of numerical data are regimes of *hyperbolic*, *adiabatic* and *tidal* encounters. We distinguish regimes of hyperbolic (near-parabolic), adiabatic and tidal encounters by plotting the curves $e' = 2$, $V/r_p = v_{cir}/a^4$ and $r_p = $ a, respectively (e.g., Heggie 2006). These curves are illustrated in Figure 1 with the diagonal green line, the blue curve and the vertical red line, correspondingly.

Encounters to the right of the vertical solid red line in Figure 1 ($r_p = a$) are tidal where $r_p \gg a$ and at larger $r_p$ the encounter is adiabatic which means that the angular speed of the planet is much larger than that of the passing star (Heggie 2006; Spurzem et al. 2009). The path of the passing star is hyperbolic with high eccentricity at very large $r_p$. If $V$ is much larger than the orbital speed of the planet then the encounter is *impulsive* and if $V$ is small compared to the orbital speed of the planet, then a tidal encounter is always adiabatic (Spurzem et al. 2009). At the bottom of the diagram, above the horizontal axis, in the tidal regime, are *near-parabolic*, adiabatic encounters (Heggie 2006) (not shown in the Figure 1). The two magenta vertical dashed lines in each panel correspond to K = 10 and K = 100, respectively (assuming that $m_1 = m_3$) measuring the strengths of encounters in which smaller K values are associated with stronger encounters (Cai et al. 2017).

The planetary system of star S1 is perturbed by three neighbouring stars; S6, S19 and S33. Planets of star S9, which has closer distance to the SMBH compared to S1, feel encounters by four nearby stars: S12, S13, S14 and S23. Whereas the planetary system of star S54 which is further away from the SMBH, undergoes weaker encounters with two stars S89 and S145 in its vicinity. At the top of the diagram are non-adiabatic, impulsive encounters, which covers most situations although more frequent for planets associated to S1 and S9. In addition, the maximum distribution of encounters shifts to a lower values of $K$ for the stars S1 and S9. This confirms that the frequency and the strength of encounters increase as the host stars reside closer to the SMBH.

---

[4] $v_{cir}^2 \sim GM_{12}/a$ is the circular velocity at radius $a$.

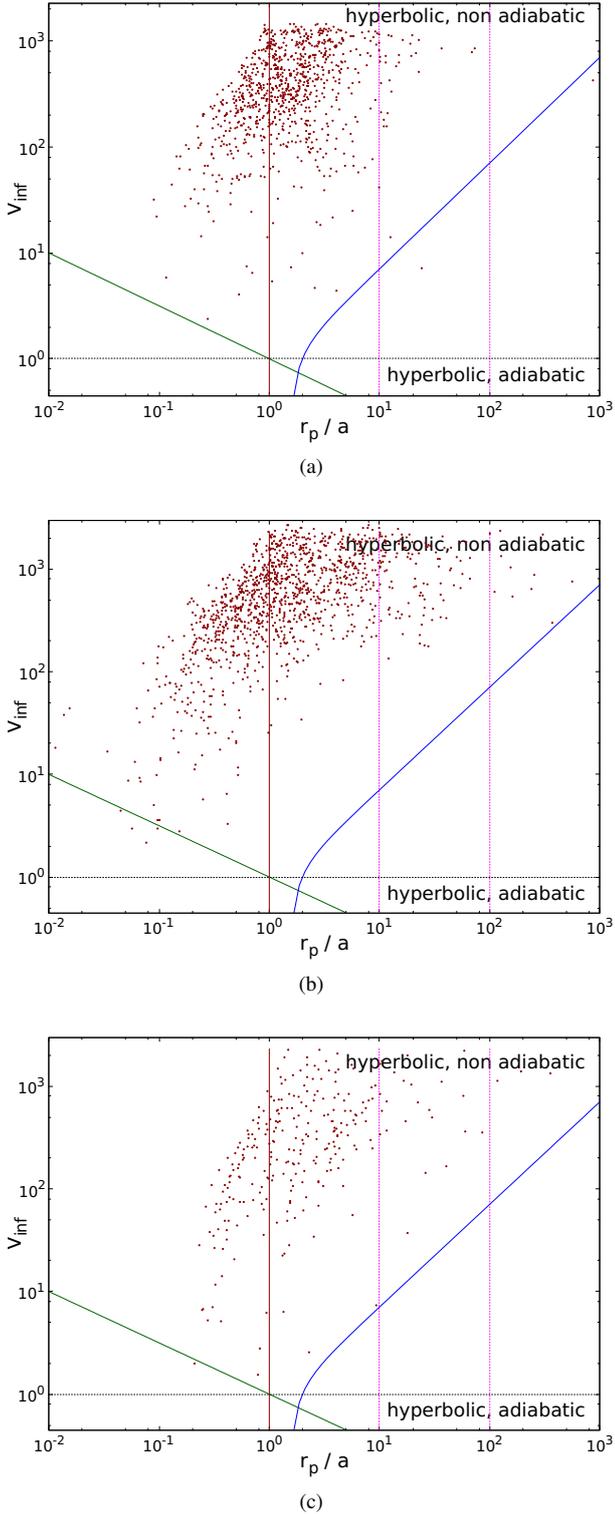

**Figure 1.** Location of encounters for planetary systems of (a) S1, (b) S9, and (c) S54, plotted according to $V_{inf}$ (the relative velocity of the perturber with respect to the host star, scaled to the average orbital speed of the outermost planet) as a function of $r_p/a$ (the minimum distance in units of the semimajor axis of the outermost planet). Above the diagonal green line encounters are hyperbolic, below this line they are nearly parabolic. The regime of adiabatic encounters (where $V \ll v_{cir}$) and impulsive encounters (where $V \gg v_{cir}$) are split with the blue curve. All details of the boundaries' definitions are given in the main text.

Compared to "hard" and "soft"[5] planetary systems of Spurzem et al. (2009) (figure 1 and figure 2), the hyperbolic, non-adiabatic regime is more destructive to the targeted planetary systems in our simulations whereas only a small number of encounters with "hard" planetary systems are non-adiabatic and there is a considerable number of near-parabolic, adiabatic encounters. In the model for the "soft" planetary systems of Spurzem et al. (2009), similar to our results, there are many non-adiabatic encounters with hyperbolic speeds and the number of near-parabolic encounters is negligible. Moreover, "soft" planets of Spurzem et al. (2009) nearly experience encounters where the $V_{inf}$ of the encountering star is larger than the orbital velocity of the planet. In the case of "hard" planets, they find a large amount of encounters with the opposite case. Looking at Figure 1, a horizontal line at $V_{inf} =1$ separates "hard" and "soft" for the outermost planet. In Cai et al. (2017), on the other hand, planetary systems face more frequent encounters in the hyperbolic, adiabatic regimes although the encounters shift downward to the near-parabolic regime in denser cluster environments.

### 3.2 Analysis of Planetary Systems Evolution

Within the time window of our simulations, we find that ∼ 51% of planets survive on stable orbits around their host S-stars even after the close interaction with the SMBH. On the other side, ∼ 49% of planets, mainly the gas giants, are captured by the SMBH and remain as *starless* planets on individual orbits around Sgr A*. The Uranus and Saturn analogues have high probability of being stripped away from their parent star due to their wide orbits and strong perturbations from Jupiter. These intrinsically unstable planets leave usually the planetary system after the first passage around the SMBH.

This work shows that the vast majority of planets from Mercury till Uranus are lost in a few thousands years for the S-stars around Sgr A*. Our simulations cover multiple pericentric passages for the S-stars with $a_* < 0.05$ pc. The shortest period is that of S2 (16 yr) and 4000 yr is ∼ 253 times the S2 period. Nevertheless, to validate the long–term stability of planetary systems bound to S–stars with $a_* \geqslant 0.05$ pc we performed some example simulations and prolonged the time from 4000 yr to 1 Myr for these samples. We noticed that the planetary systems for the stars with $a_* \geqslant 0.05$ pc remain stable around the host up to 1 Myr. With this test, the secular stability of the planets assigned to S85, with the longest period (∼ 3500 yr), is also authenticated.

Figure 2 displays the fraction of stable star-planet systems after the entire time–span of the simulation. The Jupiter–like planets around more massive stars of the early–type group have a higher chance of survival because of their vicinity to the host stars. As a general conclusion, among the planets around young S-stars, if just one remains bound to the host until the end of the simulation, is Jupiter, but if two survives are the closest to the host, i.e. Jupiter and Saturn. Mercury–like and Venus–like (short-period) planets are more resilient when around low–mass, old, S–stars.

---

[5] "Hard" and "soft" refer to two ranges of semimajor axes of planetary orbits in Spurzem et al. (2009). In their work a planetary system is considered "hard" when its semimajor axis lies within 0.03–5 AU, and "soft" when its semimajor axis is 3–50 AU.

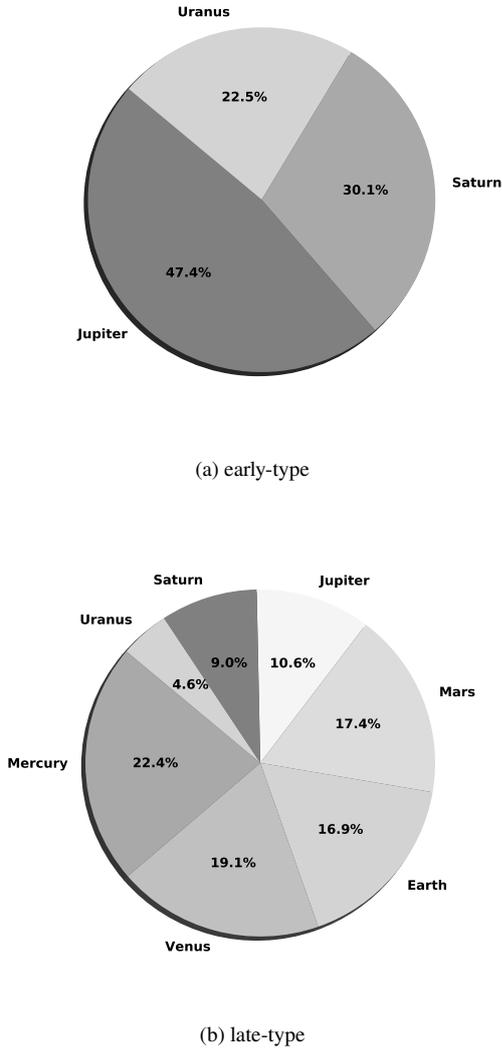

**Figure 2.** Fraction of survived planets around the early-type (upper panel a) and late-type host stars (lower panel b).

The time for an S-stars to lose entirely or partially its planetary system varies in dependence on both the vicinity to the SMBH and on the mass of the host star. Stars whose orbital semi-major axis is small, in the range 0.004pc < $a_*$ < 0.01pc, lose their planets even before completing one full orbit around the SMBH. On the other hand, S–stars with larger orbital semi-major axes, i.e. 0.01pc < $a_*$ < 0.2pc, are usually able to maintain their planetary systems intact until the end of simulation. Members of the CW, namely S66, S67, S83, S87, S91, S96, S97 and R44, preserve their planetary systems through the entire simulation.

Moreover, the larger the host mass the stronger the planetary systems are against external perturbations. As an example, the corresponding planets' vulnerability rate, at which the planets lose their stability around the host, for stars in the range 0.004 pc , $a_*$ <0.2 pc, is demonstrated in Tables 2 and 3. In this sample, the mass of early–type host stars is assumed $m_*$=10 $M_\odot$, the mass of late–type stars is supposed to be $m_*$=1 $M_\odot$; the initial inclination of the planetary system relative to star–SMBH orbital plane is $i_{in}$=20°.

We picked as examples the evolution of orbital elements for stable planets around S17 (late–type) and S1 (early–type). Figure 3 and 4 show the evolution of the orbital parameters between the host star and planets, the mutual semimajor axis ($a_{in}$), eccentricity ($e_{in}$) and inclination ($i_{in}$), of the planets around their host star. These figures represent a late–type star (S17) and an early–type star (S1) with quite small semi-major axes and stable planetary systems. The planetary system in Figure 3 initially consists of 7 Solar-type planets. The semi-major axis and eccentricity of star S17 are $a_*$ = 0.0143 pc, $e_*$ = 0.397, respectively, and for this example we assumed the mass of $m_*$ = 1.5 $M_\odot$ for the star. The initial mutual inclination of the planetary system with respect to the orbital plane of the host star is $i_{in}$=60° which is illustrated with horizontal dotted lines. Saturn and Uranus are excluded in this plot because they are almost instantly detached from the planetary system.

Figure 3 indicates the existence of eccentricity–inclination variations with different periods (one is long and the other one much shorter). To see whether or not such oscillations are in relation with the Kozai-Lidov phenomenon (Kozai 1962; Lidov 1962), the vertical dot–dashed lines in Figure 3 plots the quadrupole-order timescale of Kozai–Lidov oscillations (e.g. Perets & Naoz 2009; Antonini & Perets 2012; Antognini 2015))

$$T_{KL}^{quad} = \frac{2P_{out}^2}{3\pi P_{in}} \frac{m_1 + m_2 + m_\bullet}{m_\bullet} (1 - e_{out}^2)^{3/2}, \quad (6)$$

where $m_1$ and $m_2$ are the masses of the innermost two bodies (star and planet) orbiting each other with period $P_{in}$ and around the external perturber (SMBH) of mass $m_\bullet$ with period $P_{out}$ and and eccentricity $e_{out}$.

The five inner planets (Mercury, Venus, Earth, Mars and Jupiter) in Figure 3 remain bound around the parent star with small–amplitude oscillations in the inner planets' eccentricities and inclinations. Especially for Mercury and Venus these variations rise to a significant extent partly because of the shorter semi-major axes. The corresponding Kozai–Lidov timescale for Mercury is greater than the time of our integration as illustrated in Figure 3. For Venus, Earth and Mars analogous the Kozai–Lidov time is not in agreement with the period of the oscillations as well. For Jupiter the typical Kozai–Lidov timescale due to the SMBH is approximately matching when the excitation of the inclination (eccentricity) reaches the utmost values. The most giant planet, Jupiter, may behave as protectors of the inner ones as well. The two outermost planets, Saturn and Uranus, migrate outwards to $a \sim 50$ and $a \sim 100$ AU from the primary host star (not shown in Figure 3). However, the remaining five planets survive around their host star until the end of the simulation going through approximately 50 encounters with the SMBH. The amplitude of eccentricity oscillations does not attain the maximum value achieved through the Kozai–Lidov mechanism given by

$$e_{in,max} = \sqrt{1 - \frac{5}{3} \cos^2 i}. \quad (7)$$

Furthermore, the octupole-order Kozai–Lidov cycle operates over a timescale that is approximately given by $T_{KL}^{oct} \sim T_{KL}^{quad}/\sqrt{\epsilon_{oct}}$ where $\epsilon_{oct}$ measures the strength of the octupole-order term relative to the quadrupole-order term of the Hamiltonian (Antognini 2015)

$$\epsilon_{oct} = \frac{|m_1 - m_2|}{m_1 + m_2} \frac{a_{in}}{a_{out}} \frac{e_{out}}{1 - e_{out}^2} \quad (8)$$

so that $T_{KL}^{oct} \sim$ a few Myr for planets assigned to S17.

**Table 2.** For early–type S-stars with semimajor axis in the [0.004,0.2] pc range the table gives the time needed to detach a planet from the host star ($t_{det}$, three right columns) in units of the orbital period of the host star ($P_*$) around the SMBH. The semi-major axes ($a_*$) of the various S–stars and their orbital periods ($P_*$) are also given ($i_{in}$=20°). The term 'bound' indicates that the planet stays bound to the host star. The symbol '*' refers to stars belonging to the clockwise disc.

| | | early–type S–stars; 0.004 pc < $a_*$ <0.2 pc | | |
|---|---|---|---|---|
| Star | $a_*$ (pc) | $P_*$ (yr) | $t_{det}$ Jupiter | $t_{det}$ Saturn | $t_{det}$ Uranus |
| S55 | 0.0043 | 12.8 | 0.13 | 0.88 | 0.06 |
| S2 | 0.0050 | 16.0 | 1.95 | 0.37 | 0.10 |
| S18 | 0.0096 | 41.9 | bound | 1.07 | 0.03 |
| S5 | 0.0100 | 45.7 | 0.81 | 0.78 | 0.51 |
| S23 | 0.0102 | 45.8 | bound | 1.01 | 0.02 |
| S13 | 0.01065 | 49 | bound | 43.7 | 0.04 |
| S9 | 0.01098 | 51.3 | 49.37 | 11.80 | 0.01 |
| S14 | 0.01155 | 55.3 | 21.89 | 0.95 | 0.03 |
| S12 | 0.0120 | 58.9 | 0.98 | 1.93 | 0.37 |
| S4 | 0.0144 | 77.0 | bound | bound | 1.06 |
| S60 | 0.0156 | 87.1 | bound | 0.98 | 0.01 |
| S8 | 0.0163 | 92.9 | 7.74 | 0.96 | 0.01 |
| S175 | 0.0170 | 96.2 | 0.94 | 0.93 | 0.04 |
| S29 | 0.0170 | 101.0 | bound | 1.96 | 0.01 |
| S31 | 0.0180 | 108.0 | bound | bound | 0.07 |
| S19 | 0.0210 | 135.0 | bound | 4.04 | 0.02 |
| S1 | 0.0240 | 166.0 | bound | bound | 5.75 |
| S33 | 0.0260 | 192.0 | bound | bound | 1.03 |
| S6 | 0.0265 | 192.0 | bound | 1.0 | 0.02 |
| S42 | 0.0380 | 335.0 | bound | bound | bound |
| S71 | 0.0390 | 346.0 | bound | 0.95 | 5.0 |
| S67* | 0.0450 | 431.0 | bound | bound | bound |
| S54 | 0.0480 | 477.0 | bound | 0.72 | 0.05 |
| S22 | 0.0530 | 540.0 | bound | bound | bound |
| S83* | 0.0600 | 656.0 | bound | bound | bound |
| S96* | 0.0600 | 662.0 | bound | bound | bound |
| S66* | 0.0610 | 664.0 | bound | bound | bound |
| R34 | 0.0730 | 877.0 | bound | bound | bound |
| S91* | 0.0770 | 958.0 | bound | bound | bound |
| S97* | 0.0940 | 1270.0 | bound | bound | bound |
| S87* | 0.1100 | 1640.0 | bound | bound | bound |
| R44* | 0.1600 | 2730.0 | bound | bound | bound |

**Table 3.** Same as Table 2 but for late–type S-stars. The semimajor axis in the [0.004,0.2] pc range the table gives the time needed to detach a planet from the host star ($t_{det}$, three right columns) in units of the orbital period of the host star ($P_*$) around the SMBH. The semi-major axes ($a_*$) of the various S–stars and their orbital periods ($P_*$) are also given ($i_{in}$=20°). The term 'bound' indicates that the planet stays bound to the host star.

| | | late–type S–stars; 0.004 pc< $a_*$ <0.2 pc | | | | | | |
|---|---|---|---|---|---|---|---|---|
| Host Star | $a_*$ (pc) | $P_*$ (yr) | $t_{det}$ Mercury | $t_{det}$ Venus | $t_{det}$ Earth | $t_{det}$ Mars | $t_{det}$ Jupiter | $t_{det}$ Saturn | $t_{det}$ Uranus |
| S38 | 0.0057 | 19.2 | bound | 46.50 | 2.0 | 0.95 | 72.23 | 2.22 | 0.15 |
| S21 | 0.0088 | 37.0 | bound | bound | 4.0 | 1.93 | 0.24 | 0.23 | 0.23 |
| S17 | 0.0143 | 76.6 | bound | bound | bound | bound | 1.01 | 0.06 | 0.01 |
| S27 | 0.0183 | 112.0 | bound | 3.95 | 0.65 | 0.65 | 0.25 | 0.64 | 0.04 |
| S24 | 0.038 | 331.0 | bound | bound | bound | bound | 0.01 | 0.08 | 0.04 |
| S89 | 0.044 | 406.0 | bound | bound | bound | bound | bound | bound | 0.02 |
| S145 | 0.045 | 426.0 | bound | bound | bound | bound | bound | bound | 0.11 |
| S85 | 0.185 | 3580 | bound | bound | bound | bound | bound | bound | bound |

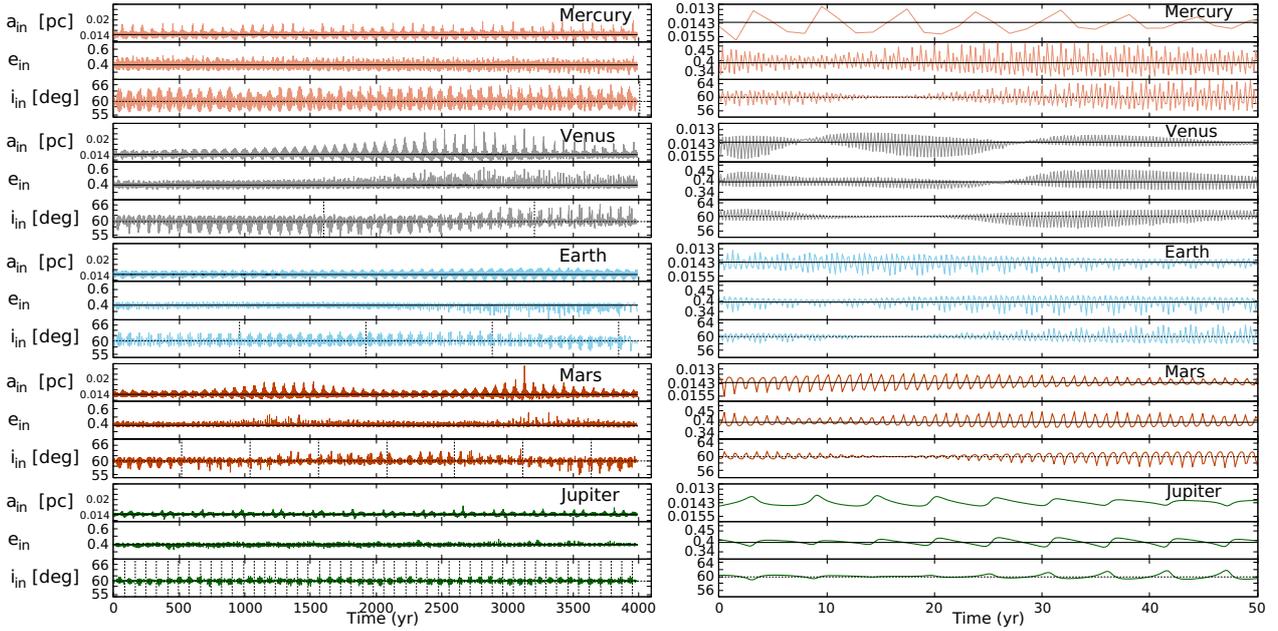

**Figure 3.** Evolution of the orbital parameters of the stable planets assigned to S17 with semi-major axis $a_* = 0.0143$ pc and eccentricity $e_* = 0.397$ (solid black line) and assumed mass of $m_* = 1.5$ M$_\odot$. The initial mutual inclination of the planetary system with respect to the orbital plane of the host star is $i_{in}=60°$ which is illustrated with horizontal dotted lines. The vertical dot–dashed lines give the Kozai–Lidov timescale due to the SMBH. Saturn and Uranus are omitted in this plot since they are detached from the planetary system. The right figure is the same simulation as the left, but shows only the first 50 yr of the integration.

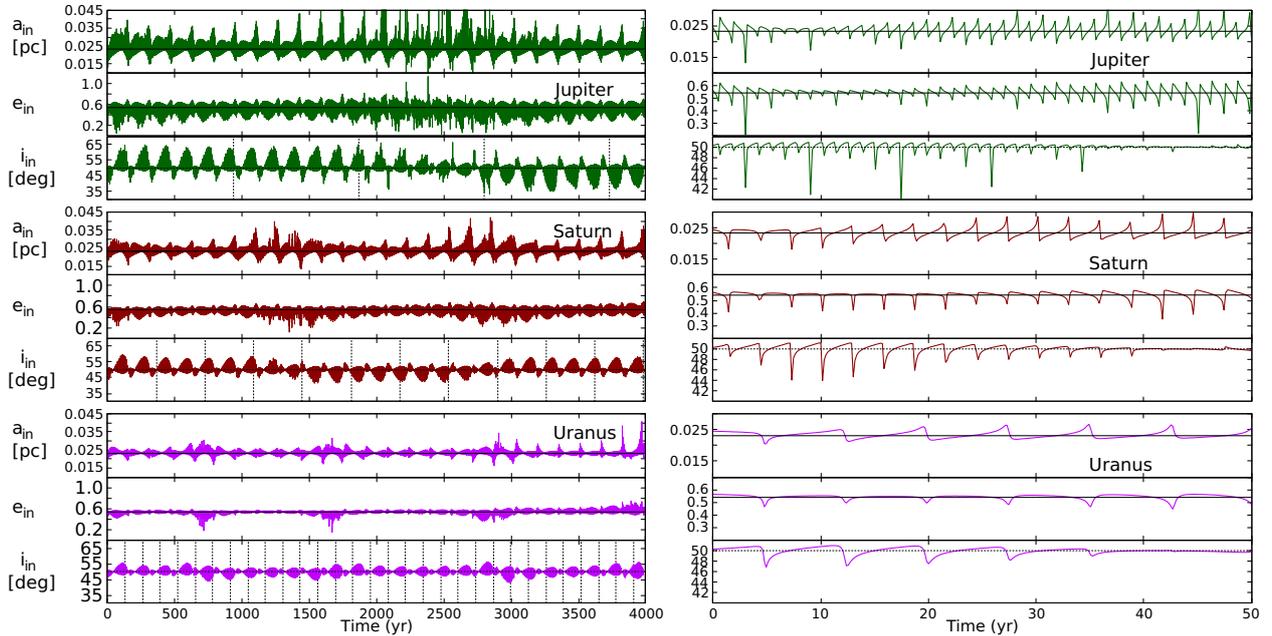

**Figure 4.** Simulation results of the evolution of the orbital parameters of the planets assigned to S1 with semi-major axis $a_* = 0.0231$ pc and eccentricity $e_* = 0.556$ (solid black line) and assumed mass of $m_* = 10$ M$_\odot$. The initial mutual inclination of the planetary system with respect to the orbital plane of the host star is $i_{in}=50°$ which is illustrated with horizontal dotted lines. The vertical dot–dashed lines give the Kozai–Lidov timescale due to the SMBH. The right figure is the same simulation as the left, but shows only the first 50 yr of the integration.

Besides, these two phenomena leads to the oscillations in eccentricity and inclination:

- short oscillations: the eccentricity reaches a maximum likely when the semi-major axis of the planet orbit is pointed toward or away from the host star (or simply when the planet has a semi-major axis smaller or larger than that of its parent star with respect to the SMBH).
- long oscillations: estimation of the period of eccentricity–inclination variations shows that there is a coupling between Earth and Mercury on a timescale of ∼ 175 days, with each showing approximately similar variations in their eccentricities and inclinations, although stronger for Mercury because of the shorter semi-major axis to the parent star. For Mars and Venus on the other hand, the effect of the 1:3 near-resonance between the two planets might lead to the eccentricity–inclination oscillations although more powerful for Venus because of closeness to the host star (similar to Mercury).

Figure 4 depicts the evolution of the orbital parameters of the planetary system orbiting around star S1 with semi-major axis $a_* = 0.0231$ pc, eccentricity $e_* = 0.556$ and assumed mass of $m_* = 10$ $M_\odot$. The initial mutual inclination of the planetary system with respect to the orbital plane of the host star is $i_{in}=50°$ which is illustrated with horizontal dotted lines as the previous example. The planetary system undergoes variations in the eccentricity and inclination. Similar to Figure 3, the vertical dot–dashed lines illustrate Kozai–Lidov timescale for the planets in this system which is not in conformity with the periodic oscillations of the inner eccentricities and inclinations. Similar to Figure 3 the short-period variations in e and i occur where the planet orbit is pointed toward or away from the host star although the the effect of the 2:1 and 4:1 near-resonances between Jupiter–Saturn and Jupiter–Uranus leads to additional long-period terms in the orbital elements. Nevertheless, going through ∼ 24 encounters with the Sgr A*, this planetary system remains relatively unperturbed and form a stable system for the rest of the simulation.

There is a small probability (∼ 0.11 %) for Jupiter and Saturn analogous around young stars to gain enough speed due to the encounter with the SMBH and trigger out of the cluster. In addition, there is also a very low probability ∼ 0.10 %) for early-type S–stars to swap their giant planets once they get very close to each other on the way of approaching the SMBH. Moreover, we estimate that the absorption probability of planets by their host star (collision of the planet onto the host star) is extremely low (close to zero).

Finally, comparing the orbital parameters of starless planets revolving around the SMBH, we find that the giant planets are consistent in semi-major axis and eccentricity with those of the G1 and G2 cloud, whereas the terrestrial-likes are not in good agreement with those of the G–clouds (see Figure 5). Likewise, Figure 6 puts in correspondence the semi-major axes and the orbital planes of the *starless* planets around the SMBH to those of the G–objects.

### 3.3 Tidal disruption of liberated planets near the SMBH

A theoretical expectation for Sgr A* as the resident SMBH of the GC is the production of tiny flares that last only a few hours but occur daily. Zubovas et al. (2012) explore the possibility that these flares could be created by disruption of small bodies, asteroids, and/or, very infrequently, by planets. In this section, we discuss fractions of liberated (starless) planets entering the tidal disruption radius of Sgr A*. The debris from tidal disruption events of these planets could also be accreted partly to the SMBH and produce flares.

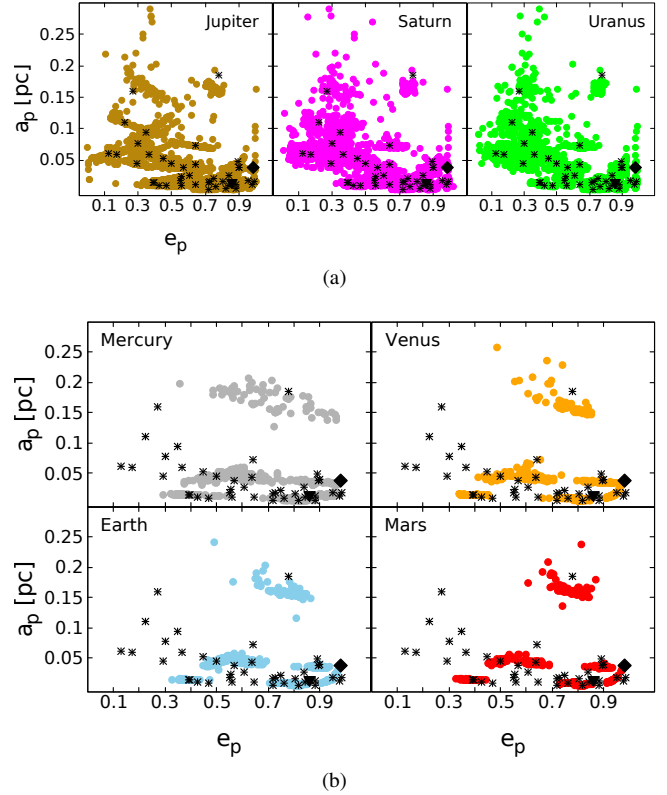

**Figure 5.** Semi-major axis ($a_p$) versus eccentricity ($e_p$) of the *starless* planets around Sgr A*. The black triangle corresponds to G1 and the black diamond represents the G2. Black stars show the distribution of the S–stars.

To compute the tidal radii, we can express the tidal radius in terms of the internal stellar structure using the standard formula (Hills 1975)

$$r_t \propto \left(\frac{6M_\bullet}{\pi\rho}\right)^{1/3}, \qquad (9)$$

where $\rho$ is the mean mass density of the object. Stars/planets of density $\rho$ are broken apart by a black hole of mass $M_\bullet$ if they pass within a distance $r_t$ of it. If Solar system analogues are abundant, our simulations proves that *starless* gas giants could undergo tidal disruption when passing close to Sgr A*. Figure 7 shows the eccentricity of starless planets versus their pericentre distances to the SMBH so that we can see which planets get close to the SMBH on their orbit and pass within the tidal disruption radius. In Figure 7, the vertical dot-dashed lines display the tidal radius for disruption of Jupiter (panel a) and Earth (panel b). From Figure 7 (b), we notice the innermost planets, do not pass within $(r_t)_{Earth}$ and the planet disruptions happen only for the three gas giants in our simulations.

In Figure 8 the fraction of tidally disrupted gas giants (Jupiter, Saturn and Uranus) as a function of the inclination orbit of planetary system with respect to stellar orbit is demonstrated. Tidal disruption fraction is maximal in the case of planetary coplanar prograde orbits ($i = 0°$) and coplanar retrograde orbits ($i = 180°$). The fraction of planet disruptions descends for inclined orbits ($0° < i < 60°$) again rises at $90°$. The minimal disruptions take place at i=60° and 120°.

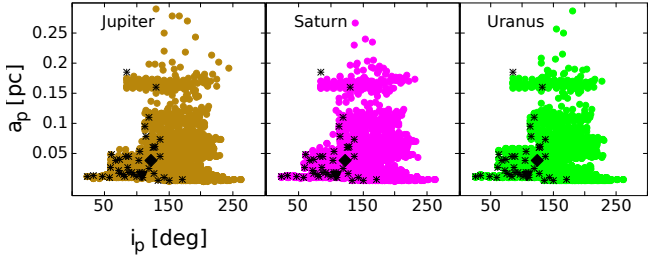

(a)

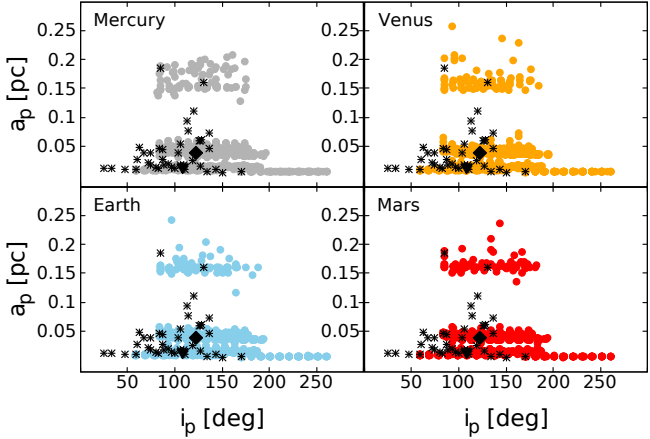

(b)

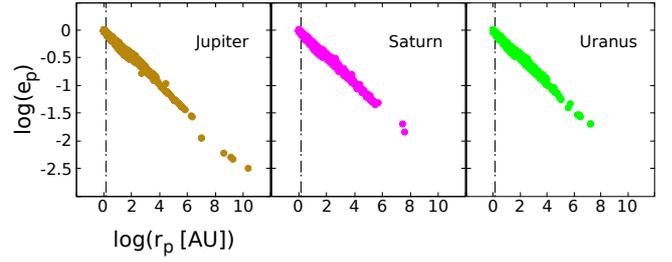

(a)

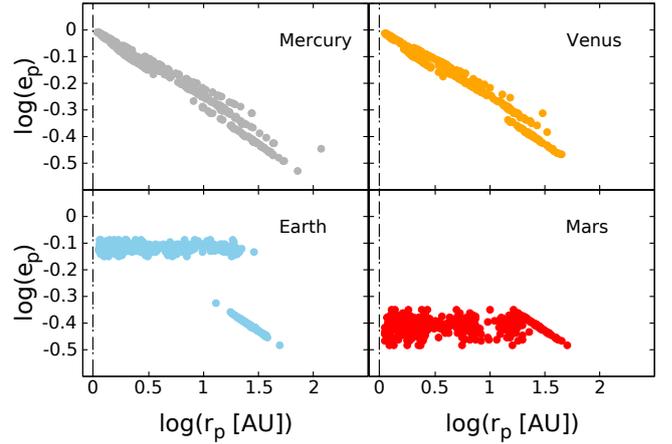

(b)

**Figure 6.** Semi-major axis ($a_p$) versus inclination ($i_p$) of the *starless* planets around Sgr A*. The black triangle corresponds to G1 and the black diamond represents the G2. Black stars show the distribution of the S–stars.

**Figure 7.** Eccentricity ($e_p$) as a function of pericentre distance $r_p$ of the *starless* planets around Sgr A*. The vertical dot-dashed lines represents the tidal disruption radii of (a) Jupiter (b) Earth by Sgr A* black hole.

## 4 SUMMARY AND DISCUSSION

In this paper we studied the short-term stability of hypothetical planetary systems (Solar-like) orbiting around the S-stars in the central arcsecond of the Galaxy. To do this, we employed a high precision, regularized *N*-body code, ARGDF, taking into account post-Newtonian corrections up to order 2.5.

We simulated 76 systems as composed of 40 S-stars each surrounded by a planetary system similar to our Solar planetary system in planets' masses, eccentricities and semi-major axes, with some limitations due to physical and computational convenience. The integration is extended up to 4000 yr to complete at least one orbit of the outermost star in the S–star cluster.

S-stars have been distinguished in 2 categories: early-type and late-type. Planetary systems assigned to late-type stars are comprised of 7 planets with same characteristics of Mercury, Venus, Earth, Mars, Jupiter, Saturn and Uranus. Planetary systems around early-type stars are made up of three giants; Jupiter, Saturn and Uranus. Our main key points can be summarized as follows:

- Planetary systems could undergo significant numbers of encounters with the nearby stars. The frequency and the strength of each stellar encounter enhance when the parent star inhabits closer to the SMBH as the maximum distribution of encounters shifts to a lower values of dimensionless parameter *K*.

- The hyperbolic, non-adiabatic encounters are more notable in our simulations while the number of near-parabolic, adiabatic encounters is negligible similar to the "soft" planetary system model of Spurzem et al. (2009).

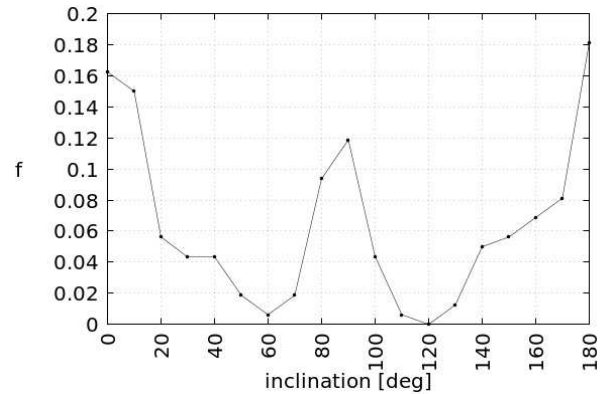

**Figure 8.** Fraction of tidally disrupted giant planets as a function of planet orbital inclination with respect to hosting star orbit.

- We determine the fraction of stable and unstable planets around the host stars and find that almost half of the planets (∼ 51%) survive around their host while the other half is indeed captured by the SMBH (∼ 49%) on randomly oriented S–like orbits similar to those of the S–stars.

- The closer the planet is to the S–star, the harder it is detached by the SMBH. The innermost planets, Mercury for late-type stars and Jupiter for early-type ones, show the maximum survival fraction around their parent stars. In harmony with the recent simulations of Vincke & Pfalzner (2018) on the evolution of compact, massive and young star clusters like Trumpler 14 and Westerlund 2 that conclude, as a result of the strong stellar interactions, planets on wide orbits are expected to be extremely rare in such environments.

- Just a minor fraction of ejected and swapped planets, ~ $10^{-3}$, is found. The ejected planets include Jupiter and Saturn around early-type stars. The swapped planets, on the other hand, are giant planets (Jupiter, Saturn and Uranus) around early-type star hosts.

- The Jupiter, Saturn and Uranus analogous create the major fraction of the *starless* planets around the SMBH.

- We discover the interesting consistency of the orbital properties of the starless planets around the SMBH with the G1 and G2 clouds in both semi-major axis and eccentricity for giant planets and in semi-major axis and inclination for both giant and terrestrial analogues.

- Our study shows that contrary to what was expected the short periodic excitation in the orbital parameters of the planets' orbits is neither due to the quadrupole-order nor to the octupole-order Kozai–Lidov mechanism. The planetary systems may be destabilised by the perturbing effect of the SMBH but the small oscillations in the mutual eccentricity and inclination of the planetary system are not produced by the Kozai–Lidov effect due to the SMBH. The planet–planet coupling and the near-resonance effect between two planets could lead to the eccentricity–inclination oscillations. However, the long-term stability of the systems is not guaranteed.

- We find that Sgr A* flares can occur due to the tidal disruption of giant planets.

Furthermore, examining the presence of planets in the proximity of the central massive black hole of our galaxy firstly requires the existence of planets around stars in the S–star cluster. The young stars in the S–star cluster have ages < 15 Myr, which means that at least some of them might be still surrounded by a protoplanetary disc rather than a fully-formed planetary system. If we take the Solar system as guidance as done here, it took considerably longer than 15 Myr to fully form. Another issue is that the systems would have already been affected in the protoplanetary disc phase. Thus the disc would have been truncated, hindering the formation of planets on Uranus- and Saturn-like orbits, possibly even Jupiter, in the first place. Nevertheless, in the recent literature (Wada et al. 2019) the possibility that a new class of planets that might form in the neighbourhood of SMBHs, potentially able to sustain Earth-like mass under certain circumstances (even 10 times the mass of the Earth), has gained attention . On the other hand, there are questions open about existence and nature of planetary systems around late–type and off main-sequence S–stars. As a matter of fact, inner terrestrial planets should have been lost during the giant branch phase (e.g., Villaver & Livio 2007; Mustill & Villaver 2012), i.e. any close-orbiting planets will be engulfed by the star and there should not be any planet within a few AU around a post-main-sequence star. At the same time, recent observational efforts by Vanderburg et al. (e.g., 2020); Jones et al. (e.g., 2021) report of the discovery of planetary companions orbiting giant and white dwarf stars. This would prove that planets could survive (very) close to their host star after leaving the main sequence. This fact intrigued us to investigate the putative Solar system as an example to determine the fate of hierarchical planetary systems of different masses and semi-major axes in the harsh environment of the Galactic Centre.

## ACKNOWLEDGEMENTS

We thank the anonymous referee for their helpful suggestions to improve the manuscript. R.S. and N.D. acknowledge the support of the DFG priority program SPP 1992 "Exploring the Diversity of Extrasolar Planets" (Sp 345/20-1). We thank Seppo Mikkola for useful discussions on the use of AR-CHAIN code. We are grateful to Manuel Arca-Sedda who made available to us the version of AR-CHAIN (ARGDF) and who generously assisted us in its use. We also thank Maryam Habibi for helpful discussions.

## DATA AVAILABILITY

The data output of this article will be shared on reasonable request to the corresponding author and is subjected to proper acknowledgement to this paper.

This paper has been typeset from a T<sub>E</sub>X/L<sup>A</sup>T<sub>E</sub>X file prepared by the author.